\documentclass[dvipsnames, twocolumn, letter]{aastex63}
\usepackage[version=4]{mhchem}
\usepackage{amsmath}

\received{May 30 2021}
\revised{August 12 2021}
\accepted{August 16 2021}
\submitjournal{ApJL}

\shorttitle{Reimagining the water snowline}
\shortauthors{Bosman \& Bergin}

\begin{document}
\title{Reimagining the water snowline }
\date{\today}

\correspondingauthor{Arthur Bosman}
\email{arbos@umich.edu}

\author[0000-0003-4001-3589]{Arthur D. Bosman}
\affiliation{Department of Astronomy, University of Michigan,
323 West Hall, 1085 S. University Avenue,
Ann Arbor, MI 48109, USA}

\author[0000-0003-4179-6394]{Edwin A. Bergin}
\affiliation{Department of Astronomy, University of Michigan,
323 West Hall, 1085 S. University Avenue,
Ann Arbor, MI 48109, USA}

\begin{abstract}
Water is a molecule that is tightly related to many facets of star and planet formation. Water's abundance and distribution, especially the location of it's snowline has thus been the subject of much study. While water is seen to be abundant in the inner region of proto-planetary disks in infrared spectroscopy, detections of water in the disk in the sub-millimeter are rare, with only one detection towards AS 205. Here we put the multitude of non-detections and the single detection into context of recent physico-chemical models. We find that the 321.2257 GHz ($10_{2,9}$--$9_{3,6}$) line detection towards AS 205 is inconsistent with a normal inner disk temperature structure and that the observed line must be masing. Furthermore, the emitting area derived from the line width, together with published analyses on water in disks around T-Tauri stars implies that the water snowline in the disk surface is at the same location as the snowline in the mid-plane. We propose that this is caused by vertical mixing continuously sequestering water from the warm surface layers into the cold disk midplane. 
\end{abstract}





\section{Introduction}
The existence of water on a terrestrial planet is one of the main requirements of life and the Earth's oceans have an enormous impact on life's origin and evolution \citep[][]{Cockell2016, Lingam2019}. The trail of water and its origin is thus one that has had strong astrophysical focus \citep[See, e.g.][ for an extensive review]{vanDishoeck2021} and this clearly extends to searching for water in the exoplanet atmospheres \citep[e.g.][]{Madhusudhan2019}.

Our current understanding suggests that the Earth formed situated within the radius of the water snowline, the transition between vapor and ice phases of water \citep[e.g.][]{Hayashi1981}. As such very little water was present on solids during Earth's assembly.  Water must have been supplied from greater distances, beyond the water snowline where icy bodies, whether planetesimals or pebbles, are found in abundance \citep{Morbidelli2000, Ida2019}. In the solar system, the water snowline location seems to have been around 2.7 au, in the middle of the asteroid belt. However, the influence of the solar system gas giants, Jupiter and Saturn could have modified this record \citep{Morbidelli2016ice,Kruijer2020}. The water snowline in the early solar system is predicted to be far closer to the Sun than 2.7 au, with values closing into 1 au, the Earths orbit \citep[e.g.][]{Mulders2015}. 

The location of the snowline is dependent on the stellar radiation field, as well as the amount of viscous heating \citep[e.g.][]{Harsono2015}. 
The location of the water snowline is therefore expected to evolve and its location directly informs on the potential supply terms of water for habitable worlds \citep{Ida2019}.

Water from within the water snowline has been observed with infrared spectroscopy \citep[e.g.][]{Carr2008, Pontoppidan2014PPVI}. \textit{Spitzer}-IRS observations show that water is abundant in the inner region of proto-planetary disks, with an emitting region that has a 1--3 au radius based on the LTE line modeling \citep{Pontoppidan2010, Carr2011, Salyk2011}.  This is confirmed by velocity resolved mid-infrared spectroscopy \citep{Pontoppidan2010res, Najita2018, Salyk2019}. The \ce{H2O} lines in the mid-infrared have high upperlevel energies, $>$ 3000 K for the velocity resolved lines and thus only trace hot gas. It is not directly clear if the extend of the emitting area is set by the abundance structure of \ce{H2O} or the temperature in the inner disk. As such these emitting area's cannot be directly used to infer the \ce{H2O} snowline. 
This is further complicated by the fact that these lines only trace the upper atmosphere as the mid-plane is obscured by optically thick dust. Any snowline inferred would thus be the surface snowline.

\citet{Blevins2016} have attempted to break this degeneracy for four well studied disks, using lower upper level energy water lines observed by \textit{Hershell} between 75 and 180 $\mu$m to probe the colder ($T >$ 300 K) regions of the disk surface. They find that the \ce{H2O} abundance does strongly drop, even in the surface layers of the disk where \ce{H2O} should not freeze-out. The radius at which the \ce{H2O} abundance drops, is inferred to be larger than emitting radii inferred from the infrared lines by a factor $\sim$2 for three out of the four sources\footnote{The fourth sources, RNO 90, has an inferred abundant \ce{H2O} radius $\sim$8 $\times$ the area necessary for the IR lines}. Implying that both excitation and abundance are important in setting infrared \ce{H2O} line strength. 

Sub-millimeter observations with ALMA are in theory perfect to supplement the existing observations. The orders of magnitude lower dust opacity at sub-millimeter wavelengths as well as the availability of \ce{H2{}^{18}O} lines allow the observations to probe deep into the disk and the low upper level energies ($<500$ K) are more sensitive to gas around the mid-plane water snowline \citep[e.g.][]{Notsu2018}. The search for water with ALMA has so far mostly led to upper limits, even in warmer and younger disks \citep[][and sec. ~\ref{sec:upplim}]{Notsu2019, Harsono2020}.  

Only one detection of water from the inner disk reservoir has been made with ALMA towards solar analogue AS 205 \citep[0.87 M$_\odot$;][]{Carr2018}. They report a detection of the 321.2257 GHz o-\ce{H2O} $(10_{2,9}$--$9_{3,6})$ water line. This line has an upper level energy of 1861 K and an Einstein A coefficient of 5.048$\times10^{-6}$. Interestingly the lower energy 322.4652 GHz p-\ce{H2{}^{18}O} $(5_{1,5}$--$4_{2,2})$ line, with an upper level energy of 468 K and Einstein A coefficient of 1.045$\times 10^{-5}$, was not detected. In this letter we will interpret the ALMA water detection and non-detections and discuss the implications for the structure of the water reservoir around the water snowline. 

\section{Water emission from within the midplane snowline}

\begin{figure}
    \centering
    \includegraphics[width = \hsize]{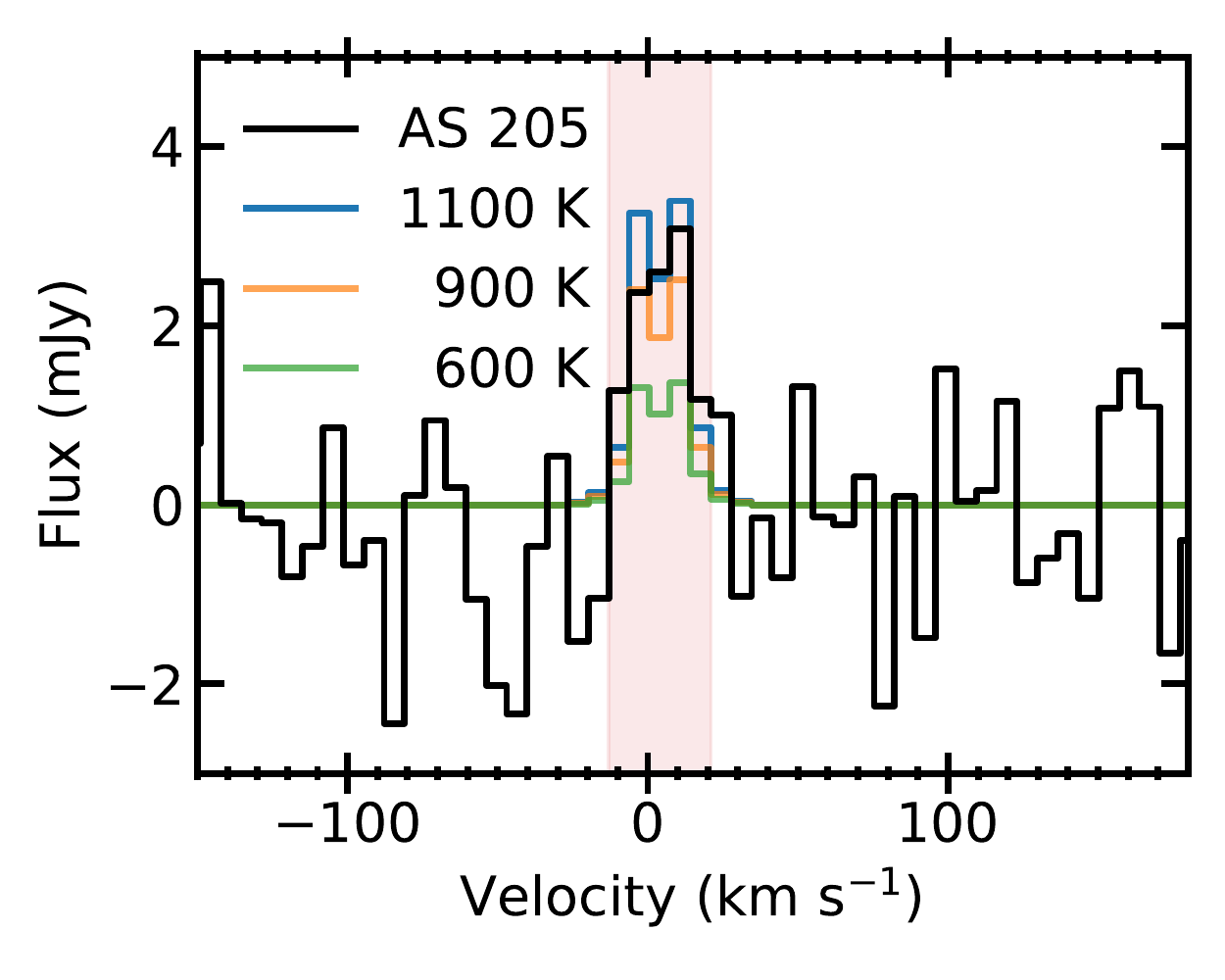}
    \caption{$10_{2,9}$--$9_{3,6}$ water line observed towards AS 205 compared to power-law intensity models with an extend of 1.8 au. Labels denote the temperature assumed at 1.8 au, and the temperature behaves as $T(R) = T_{\mathrm{1.8\,au}} (R/1.8)^{-0.5}$. Very high temperatures at 1.8 au are necessary to produce the observed line. }
    \label{fig:water_maser}
\end{figure}

We rereduced the AS 205 data from project 2016.1.00549.S (PI: Carr, J.)  using \rm{CASA} version 5.6.1. After the standard pipeline calibrations\footnote{This was done with the pipeline of \rm{CASA} version 4.7.2, for which the pipeline script was developed.} we performed two rounds of phase only self-calibration and one round of amplitude self-calibrations, giving a S/N increase on the continuum of a factor 14. The lines were then imaged using briggs weighting with a robust of 0.5, binning to a 6.8 km s$^{-1}$ channel width following \citep{Carr2018}. We do not perform any clean iterations. The spectrum for the 321.226 GHz line extracted on the peak position of the continuum is shown in Fig.~\ref{fig:water_maser}. The integrated line flux between -14 and 22 km s$^{-1}$ is 71$\pm$17 mJy km s$^{-1}$, consistent with the values from \citep{Carr2018}. We ran a couple of test on the imaging and measure similar line fluxes from the product data, when imaging the measurement set before self-calibration, when using a different channel binning (5 km s$^{-1}$ binning or a 3.4 km s$^{-1}$ velocity shift) as well as when imaging with natural weighting. This implies that this is not a feature introduced by any of the steps in data reduction. 

\begin{figure*}
    \centering
    \includegraphics[width = \textwidth]{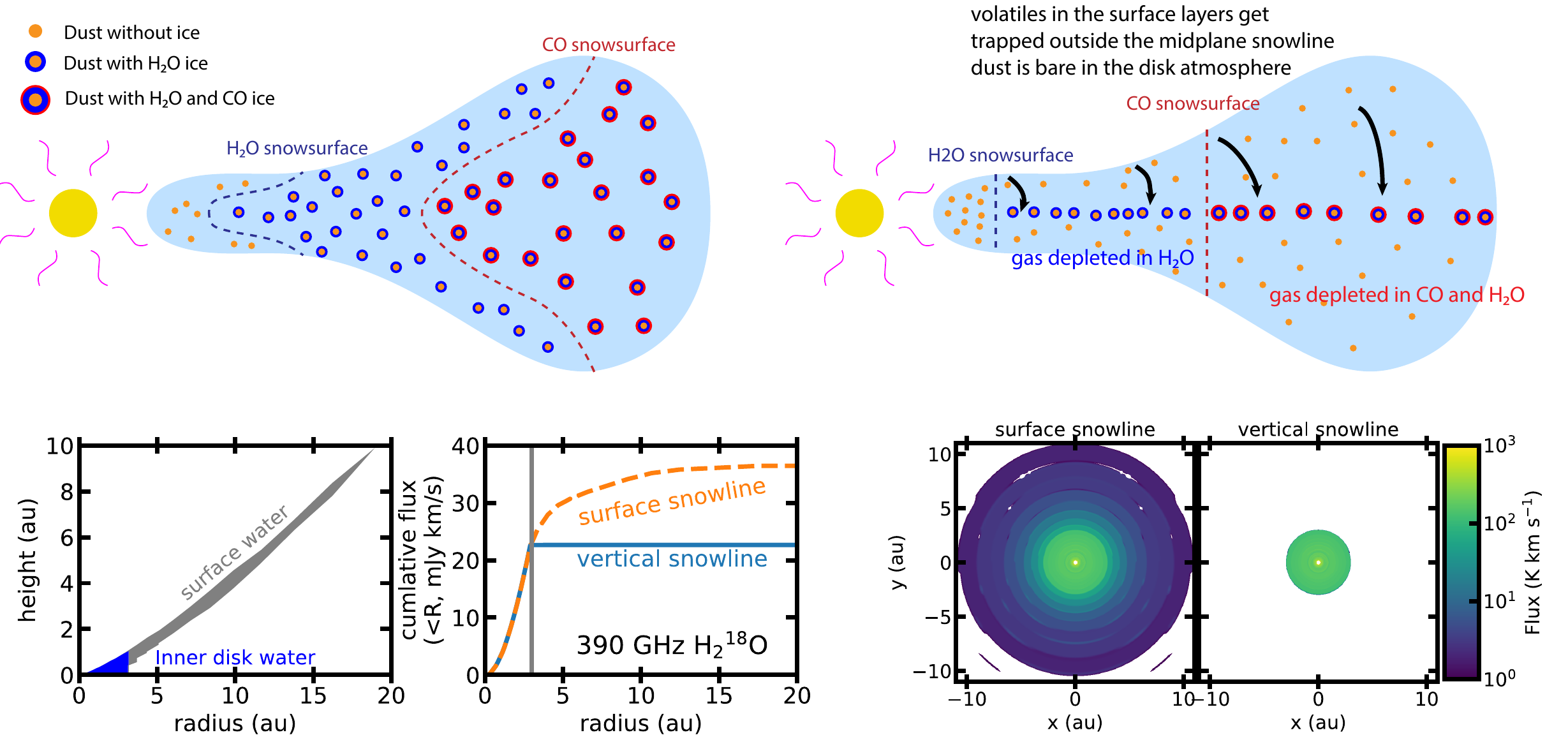}
    \caption{\label{fig:Schem_snowlines} [Top:] Schematic of the classical picture of the major \ce{H2O} and \ce{CO} snowsurfaces in a disk (left) and the behavior derived from observations \citep{Blevins2016, Du2017, Zhang2019} and predicted by models \citep{Krijt2016Water, Krijt2020}. [Bottom:] Model predictions for water emission. Left panels show the gaseous water containing regions in a AS 205 disk model \citep{Bruderer2015} with T$_\mathrm{dust} > 150 K$ and $A_V> 1$, split into the inner disk and surface reservoirs is shown on the left. The water abundance in the inner disk is assumed to be $10^{-4}$ while the surface layer water abundance is assumed to be $10^{-8}$ or $10^{-5}$ representing situations with a vertical and surface snowline. Rest of the panels show emission predictions for the $4_{1, 4}$--$3_{2, 1}$ \ce{H2{}^{18}O} line for these two models both as cumulative radial profiles as well as integrated line intensity.}

\end{figure*}

The line width, $\sim$20~km\,s$^{-1}$, implies a small water emitting area especially considering the near face-on (20 degree) inclination of AS 205. \citet{Carr2018} find the emitting area to be constrained within a 1.8 au radius from of the line kinematics. This agrees with the emitting area extracted from line kinematics and LTE modeling of mid-infrared lines origination from the AS 205 disk \citep{Salyk2011, Najita2018}. We use this emitting area to extract the emission temperature of the water line. We assume a power-law temperature profile between 0.1 and 1.8 au with a power-law coefficient $q = -0.5$. For simplicity we will assume that the line is optically thick and that the line width is given by the kinetic temperature at 1.8 au. The temperature at 1.8 au is varied to match the observed line strength. 

Figure~\ref{fig:water_maser} shows the comparison between the observed water line and the power law models. Very high temperatures ($>$ 900 K) at 1.8 au are necessary to reproduce the line flux. This temperature is significantly higher than previous temperatures derived from the inner 2 au for AS 205. Modelling of the \textit{Spitzer} water lines implies a temperature of 450 K \citep{Salyk2011}, which is consistent with the rotational temperature of the ro-vibrational \ce{^{13}CO} lines that have the same line width as the $10_{2,9}$--$9_{3,6}$ water line \citep{SalykCO2011, Banzatti2017}. Finally, \citet{Najita2018} find a temperature of 680 K to fit the velocity resolved 12.5 $\mu$m water lines also coming from within 2 au. The emission temperature necessary for the $10_{2,9}$--$9_{3,6}$ water line is thus anomalously high. Especially when considering that all the other observations trace high Einstein A coefficient lines with upper level energies $> $3000 K, significantly higher than the $10_{2,9}$--$9_{3,6}$ line upper level energy of $\sim$1800 K. The other tracers should thus be tracing hotter gas. 

The uncertainties in the observation allows for a larger emitting area, \citet{Carr2018} quote a 1$\sigma$ range of 1.1-2.5 au.  A larger emitting area would result in significantly lower emission temperatures required to reproduce the line. The $10_{2,9}$--$9_{3,6}$ line would still have to be optically thick, indicating a column of $>${}$10^{19}$ cm$^{-2}$. At gas temperatures above 300 K, this gas would also contribute to the \textit{Spitzer} spectra \citep[e.g.][]{Salyk2011}.  If temperatures are below 300 K (i.e. much larger emitting area) the lower upper level energy \ce{H2{}^{18}O} lines have similar optical depths and should have been detected.  They are not (see Sec.~\ref{sec:upplim}).

\begin{figure*}
    \centering
    \includegraphics[width = \hsize]{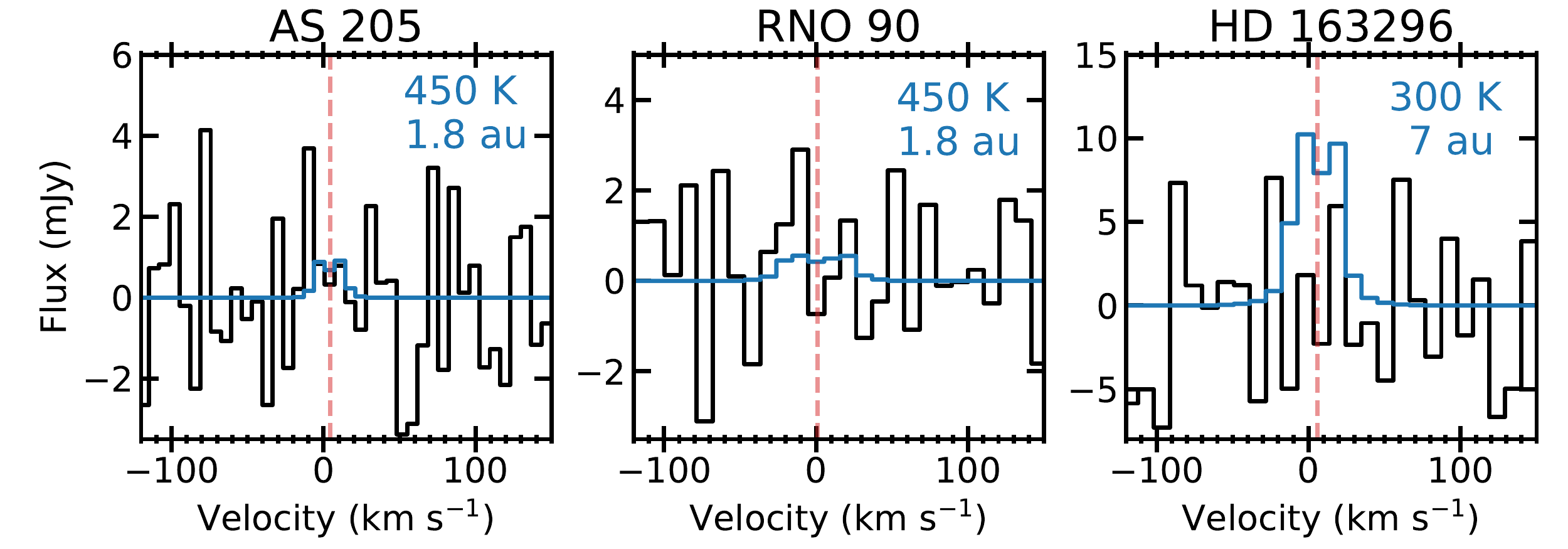}
    \caption{Non-detections of \ce{H2{}^{18}O} lines towards AS 205 N (322.465 GHz, $5_{1,5}$--$4_{2, 2}$, left), and RNO 90 and HD 163296 (390.608 GHz, $4_{1, 4}$--$3_{2, 1}$, middle and right respectively. Red vertical lines show the systemic velocity for the three systems. The blue line shows an power-law intensity model. For AS 205 and RNO 90 temperature and extent are chosen in rough correspondence to the temperatures and emitting areas derived from mid-infrared line modelling. In the case of HD 163296, the extent is given by the water snowline in the model from \citet{Notsu2016} with the temperature the difference between the surface layer gas temperature and the mid-plane dust temperature at the mid-plane snowline location.}
    \label{fig:water_non}
\end{figure*}

Assuming the same emitting region for the sub-millimeter and infrared lines, the high line strength of the $10_{2,9}$--$9_{3,6}$ line compared to the infrared lines can only be explained by assuming that this line is currently masing. The $10_{2,9}$--$9_{3,6}$ water line is often seen to be masing in star-forming regions where it traces high density shocks \citep[e.g.][]{Patel2007}, as well as in the outflows of (post-)asymptotic giant branch stars \citep[e.g.][]{Gray2016}. In these environments this line show masing behavior for \ce{H2} densities below $10^{10}$ cm$^{-3}$ \citep{Neufeld1991, Gray2016}. 

It is very difficult to extract physical properties of the line emitting regions from maser lines, especially if there is only one \citep{Neufeld1991}. As such using this line for abundance determinations in the inner disk is going to be virtually impossible. {However, the measured brightness does require a saturated maser, which puts a lower limit to the water column. }
{For densities of $10^{9}$ cm$^{-3}$ the water maser saturates around a column of $10^{18}$ cm$^{-2}$ \citep{RADEX, Daniel2011}. } 
{Assuming a water abundance of $10^{-4}$, this results in a \ce{H2} column density of $\sim${}$10^{22}$ cm$^{-2}$. The density in the \citet{Bruderer2015} model reaches $\sim${}$5\times 10^{9}$ cm$^{-3}$ at this column. This on the high side for this line to be masing, but the density depends strongly of the assumed disk structure. Furthermore a water abundance elevated above $10^{-4}$, for example due to drift \citep[e.g.][]{Ciesla2006}, would make it easier to get the required water column at a low enough density for maser activation.}

The maser boost, does make the line easier to detect. A thermal line with a 450 K emitting temperature and the same emitting region would not have been detected with 30 minutes of integration with ALMA. The maser boost allows us to use ALMA to probe the region inwards of the water snowline, and, using kinematical analysis allow us toT determine the emitting region of the water lines observed in the infrared with \textit{Spitzer-IRS} and \textit{JWST-MIRI} providing an alternative to the challenging ground-based velocity resolved mid-infrared water observations \citep{Najita2018, Salyk2019}.

\begin{table*}[!thb]
    \centering
    \caption{Source properties}
    \begin{tabular}{l c c c c c r}
    \hline \hline
        Source &  Stellar mass & Stellar luminosity & Distance & Inclination  & Snowline location & references\\
               &  ($M_\odot$) &  ($L_\odot$) & (pc) & (deg) & (au) \\
    \hline
        AS 205 & 0.871 & 2 & 128 & 20 & 1.8 & (1,2)\\
        RNO 90 & 1.5 & 4 & 125 & 37 & 1.5-11 & (3,4,5)\\ 
        HD 163296 & 2.0 & 17 & 101 & 46 & 7 & (1,6) \\
    \hline
    \end{tabular}
    \tablecomments{References: 1 \citet{Andrews2018}, 2 \citet{Carr2018}, 3 \citet{Salyk2011}, 4 \citet{Pontoppidan2011} 5 \citet{Blevins2016}, 6 \citet{Notsu2019}}
    \label{tab:source_prop}
\end{table*}

\section{Reimagining the water snowline}
The small water emitting area for the $10_{2,9}$--$9_{3,6}$ combined with a similar emitting area for many other water lines \citep{Salyk2011, Najita2018, Carr2018} implies that the water emitting area is strongly confined within the AS 205 disk. The non-detection of the \ce{H2{}^{18}O} lines further supports this and implying a small water emitting radius \citep[][, Sec.~\ref{sec:upplim}]{Carr2018, Notsu2019}. In an AS 205 specific model, \citet{Bruderer2015} find that at $\sim$ 2 au, the midplane temperature is around 200 K, the expected temperature of the \ce{H2O} snowline \citep[assuming a density of $10^{16}$ cm$^{-3}$,][]{Harsono2015}. 
For AS 205, the physical extent of the surface water reservoir thus seems to be as big as the mid-plane water reservoir. This is in conflict with the classical picture of the water snowsurface (see Fig.~\ref{fig:Schem_snowlines}). This is inline with the observational results that consistently find a strong jump in the water abundance in the surface layers outside of the midplane snowline location \citep{Meijerink2009, Bergin2010, Hogerheijde2011, Blevins2016}. 

This has previously been inferred as a drop of the water abundance due to inefficient \ce{H2O} formation in colder ($<$300 K) gas \citep{Blevins2016}. However, not sequestering the elemental oxygen in water, would imply a very high \ce{CO2} abundance ($\sim10^{-4}$ with respect to \ce{H2}), which is strongly in conflict with current observations \citep{Pontoppidan2014, Bosman2017, Bosman2018}. Instead we propose that this drop in the water abundance is due to chemo-dynamical processing, analogues to what happens in the outer disk with CO outside of the CO mid-plane snowline \citep{Kama2016, Krijt2018, Krijt2020}. 

In the outer disk CO is thought to be depleted by a combination of vertical mixing, dust settling and chemical processing. The vertical mixing moves gaseous CO from above the surface snowline to the mid-plane where it gets stuck on large settled grains, while chemical conversion transforms CO into less volatile species (\ce{CO2}, \ce{H2O}, \ce{CH4}) which also get stuck on the grains. This leads to an CO abundances that is 1-2 orders of magnitude lower in the disk surface.  {This is below the expected surface layer abundance even in gas and dust layers that are above the CO sublimation temperature}. This can be clearly see in the CO abundance profiles from \citet[][, Fig. 8]{Zhang2019}, which have as ISM CO abundance within the snowline, but a strongly depleted CO abundance outside, which implies the CO abundance lowers over the entire vertical extend of the disk. On a global scale, there thus is only a vertical snowline at the mid-plane snowline radius and no CO snow surface. This is shown schematically in Fig.~\ref{fig:Schem_snowlines}. 

To explain the observed water abundance structures we invoke a similar effect near the \ce{H2O} snowline. \citep{Krijt2016Water} shows that in an iso-thermal column vertical mixing and dust growth can lead to strong water depletion above the water snow surface. This would naturally explain the strong jump in water abundance at the mid-plane snowline location. 

This process would even help explain the relatively low \ce{CO2} mid-infrared fluxes observed towards proto-planetary disks \citep{Pontoppidan2014, Woitke2019}. In the disk surface, \ce{CO2} that is present will mostly be converted into \ce{H2O} and \ce{CO} by the strong UV field and modest temperatures (100-300 K) in the disk surface between the \ce{H2O} and \ce{CO2} icelines \citep[e.g.][]{Bosman2018}. If the \ce{H2O} continuously gets sequestered into the mid-plane, this would remove the oxygen necessary for \ce{CO2} formation. This leaves the gas with only \ce{CO} carrying a significant fraction of the gaseous oxygen budget, leading to a gas-phase C/O of $\sim$ 1 in the disk surface between the \ce{H2O} and \ce{CO} mid-plane icelines. 

\section{\ce{H2{}^{18}O} upper limits}
\label{sec:upplim}

To see if our picture of a vertical water snowline holds, we looked at published and archival ALMA data of \ce{H2O} lines with lower upper level energies than the $10_{2,9}$--$9_{3,6}$ water line which should better trace the 150-300 K water vapor. This encompasses two $(5_{1,5}$--$4_{2,2})$ \ce{H2{}^{18}O} non-detections \citep{Carr2018, Notsu2019} towards AS 205 and HD 163296 as well as two previously unpublished 390.608 GHz, $4_{1, 4}$--$3_{2, 1}$ \ce{H2{}^{18}O}, $E_\mathrm{up} = 322$ K, non-detections towards RNO 90 and HD 163296 [in 100 and 40 minutes with ALMA respectively, 2015.1.00847.S, PI Du, F]. Figure~\ref{fig:water_non} shows the non-detection of the $4_{1, 4}$--$3_{2, 1}$ \ce{H2{}^{18}O} line in the product data towards both disks, together with the non-detection of the $5_{1,5}$--$4_{2, 2}$ \ce{H2{}^{18}O} water line towards AS 205 \citep{Carr2018}. To put these line observations in context, Table~\ref{tab:source_prop} summarizes the some source properties for these systems. 

The non-detections towards T-Tauri stars AS 205 and RNO 90 are consistent with a small snowline radius (taken to be 1.8 au following the other results for AS 205) and a gas-dust temperature contrast of 450 K at the \ce{H2O} snowline, following the \ce{H2O} and \ce{^{13}CO} ro-vibrational temperature derived towards AS 205 at this radius \citep{Salyk2011, SalykCO2011}. 
We find a 3$\sigma$ flux upper limit between -14 and 22 km\,s$^{-1}$ of 84 mJy\,km\,s$^{-1}$ for AS 205, and a 3$\sigma$ flux upper limit between -17 and 19 km\,s$^{-1}$ of 90 mJy\,km\,s$^{-1}$ for RNO 90. 

No mid-infrared water lines have been detected towards Herbig Ae star HD 163296, as is typical for these systems \citep{Pontoppidan2010}. As such no mid-infrared derived snowline radius is available. Based on thermo-chemical modelling, \citet{Notsu2016} estimate a water snowline radius of 7 au. We make a simple prediction for the line flux for HD 163296, taking a low estimate for the gas temperature in the surface layer, 300 K at the 7 au snowline radius. This results in a model line flux of 380 mJy km s$^{-1}$, compared to the 500 mJy km s$^{-1}$ prediction of a Herbig Ae disk \citep{Notsu2018}. This would have been detected at $>$ 3$\sigma$  which correspond to 290 mJy\,km\,s$^{-1}$ between -13 and 23 km\,s${-1}$. This is consistent with the non-detection of the water lines around 322 GHz \citep{Notsu2019} and implies either a smaller emitting region, or optically thick dust at 300-400 GHz in the warm molecular layer suppressing water line emission. 

\section{Summary}
We have re-examined the detection of the water line at 321.2256 GHz towards AS 205 N from \citet{Carr2018} and in light of {water emission line} constraints on the inner disk of AS 205 N conclude that this line is a maser originating from the water reservoir within the mid-plane snowline of the AS 205 N. Furthermore, we propose that vertical mixing and dust settling have a strong impact on the abundance of water above the surface snowline, creating a jump profile in the water abundance at the location of the mid-plane snowline, with a low \ce{H2O} abundance at outside the water snowline. Warm water emission thus always traces the emitting area within the water mid-plane snowline. This explains the lack of \ce{H2{}^{18}O} water line detections with ALMA to date as well as suggests we can use masing \ce{H2{}^{16}O} transitions such as the line at 321.2256 GHz to kinematically probe the water snowline location down to radii that ALMA would otherwise not be sensitive to.

\acknowledgments 
This paper makes use of the following ALMA data: ADS/JAO.ALMA\#2015.1.00847.S, ADS/JAO.ALMA\#2016.1.00549.S. ALMA is a partnership of ESO (representing its member states), NSF (USA) and NINS (Japan), together with NRC (Canada), MOST and ASIAA (Taiwan), and KASI (Republic of Korea), in cooperation with the Republic of Chile. The Joint ALMA Observatory is operated by ESO, AUI/NRAO and NAOJ. The National Radio Astronomy Observatory is a facility of the National Science Foundation operated under cooperative agreement by Associated Universities, Inc. ADB and EAB acknowledge support from NSF Grant\#1907653 and NASA grant XRP 80NSSC20K0259. 
\software{Astropy \citep{astropy2013,astropy2018}, SciPy \citep{Virtanen2020},  NumPy \citep{van2011numpy}, Matplotlib \citep{Hunter2007}, RADEX \citep{RADEX}.}

\bibliographystyle{aa.bst}
\bibliography{Lit_list}

\end{document}